\documentclass[12pt]{iopart}     

\usepackage{graphicx}
\usepackage{bm}
\usepackage{amsfonts}

\begin{document}

\title{Asymmetric transmission: a generic property of lossy periodic interfaces}

\author{E Plum, V A Fedotov and N I Zheludev\footnote{www.metamaterials.org.uk}} \address{Optoelectronics Research Centre, University of Southampton, SO17 1BJ,
UK} \eads{\mailto{erp@orc.soton.ac.uk},
\mailto{niz@orc.soton.ac.uk}}

\date{\today}

\begin{abstract}
Asymmetric transmission of circularly polarized waves is a
well-established property of lossy, anisotropic, two-dimensionally
chiral patterns. Here we show that asymmetric transmission can be
observed for oblique incidence onto any lossy periodically structured plane. Our results greatly expand the range of natural and artificial materials in which directionally asymmetric transmission can be expected making it a cornerstone electromagnetic effect rather than a curiosity of planar chiral metamaterials.  Prime candidates for asymmetric transmission at oblique incidence are rectangular arrays of plasmonic spheres or semiconductor quantum dots, lossy double-periodic gratings and planar metamaterial structures.
\end{abstract}




\maketitle

Since Hecht and Barron \cite{HechtBarron}, and Arnaut and Davis
\cite{Arnaut} first introduced planar chiral structures to
electromagnetic research they have become the subject of intense
theoretical and experimental investigations with respect to the
polarization properties of scattered fields in both linear and
nonlinear regimes
\cite{PRE_Prosvirnin_2003_chiralDiffraction,Bedeaux,PRL_Papakostas_2003_2dChirality,Schwanecke_2003,PRL_KuwataGonokami_2005_GiantOpticalActivity,
PRA_2DonSubstrateTheory,KuwataWaveguide,OptLett_Decker_2007_CircDichroismAlignedRosettes,L_NanoParticleSHG}.
It was understood that planar chirality is essentially different in
symmetry from 3D chirality and should lead to a new fundamental
polarization phenomenon, the sense of which reverses for opposite
directions of light propagation. Recently such a phenomenon has
indeed been discovered in the form of asymmetric transmission, which
manifests itself as a difference in both normal incidence
transmission and retardation of circularly polarized waves incident
on opposite sides of a planar chiral structure. It was observed in
regular sub-wavelength arrays of anisotropic intrinsically 2D-chiral
meta-molecules in the microwave
\cite{PRL_Fedotov_2006_AsymmetricTransmissionMW,
APL_Plum_2DchiralASR}, terahertz \cite{PRB_Singh_2009_AsymTrans} and
optical \cite{NanoLett_Schwanecke_2008_AsymTrans} parts of the
spectrum as well as isolated plasmonic nano-structures
\cite{OptExp_Ebbesen_2008_AsymTrans} and has been linked to
directionally asymmetric absorption losses
\cite{APL_Plum_2DchiralASR, OL_Zhukovsky}.

In this Letter, we demonstrate that asymmetric transmission can be observed at \emph{any} lossy periodically structured interface. We show that structural 2D chirality \cite{JOPA_Potts_2004_chiralityModel} - which causes the effect - can arise from oblique incidence onto any periodic pattern.

\begin{figure}[t!] \centering
\includegraphics[width=120mm]{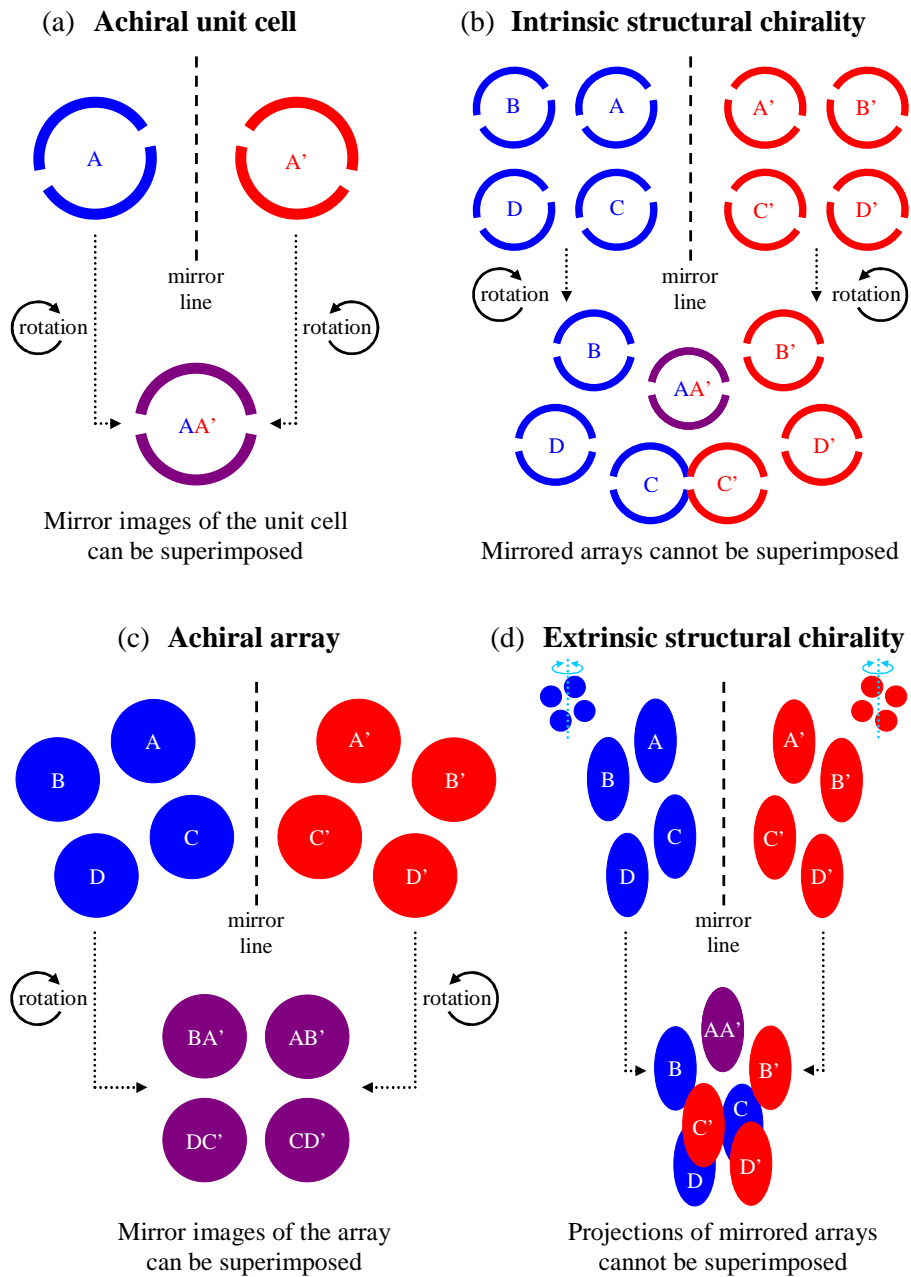}
\caption{\label{fig-struct-chirality} \textbf{Intrinsic structural
chirality:} Anisotropic achiral meta-molecules (a) can form a
structurally planar chiral array (b). While a single meta-molecule
(blue) and its mirror image (red) can be superimposed by translation
and rotation (purple), certain arrays of such meta-molecules show
planar enantiomorphism: congruency can only be achieved for one
meta-molecule in the array (purple) while the rest of the mirrored
arrays does not coincide. \textbf{Extrinsic structural chirality:}
An array (c) of highly symmetric meta-molecules (blue) is congruent
with its mirror image (red) and therefore  does not have intrinsic
chirality. However (d), when it is tilted with respect of the
observation direction its projection onto the plane normal to this
direction becomes planar chiral.}
\end{figure}

Like electromagnetic effects of molecular chirality
\cite{JOPA_Plum_2009_Extr2Dchir}, electromagnetic manifestations of
structural chirality can arise in two ways: intrinsically or
extrinsically. As illustrated in
figures~\ref{fig-struct-chirality}(a) and (b), the intrinsic form of
structural 2D chirality results from the orientation of achiral
meta-molecules placed in a planar regular array, when the lines of
mirror symmetry of the molecules and the mirror lines associated
with the array's lattice do not coincide. Note that in this case
mirror-forms of the array cannot be superimposed by translations and
rotations in the plane, which makes the entire structure 2D-chiral.

As shown by figures~\ref{fig-struct-chirality}(c) and (d),
structural 2D chirality can also be imposed extrinsically even in a
regular array containing meta-molecules of the highest symmetry.
This is achieved by tilting the array around any in-plane axis that
does not coincide with one of the array's lines of mirror symmetry.
It is easy to see that in this case the metamaterial's projection
onto the plane normal to the incidence direction becomes
structurally 2D-chiral and anisotropic. Note that chirality here
arises from the arrangement of the meta-molecules, rather than their
internal structure. This implies that various 2D-chiral phenomena
may be expected at any planar regular array containing identical
particles of any symmetry. In particular asymmetric transmission,
which has been previously perceived as an exotic effect specific to
metamaterials, may in fact be a common phenomenon.

\begin{figure*}[t!] \centering
\includegraphics[width=120mm]{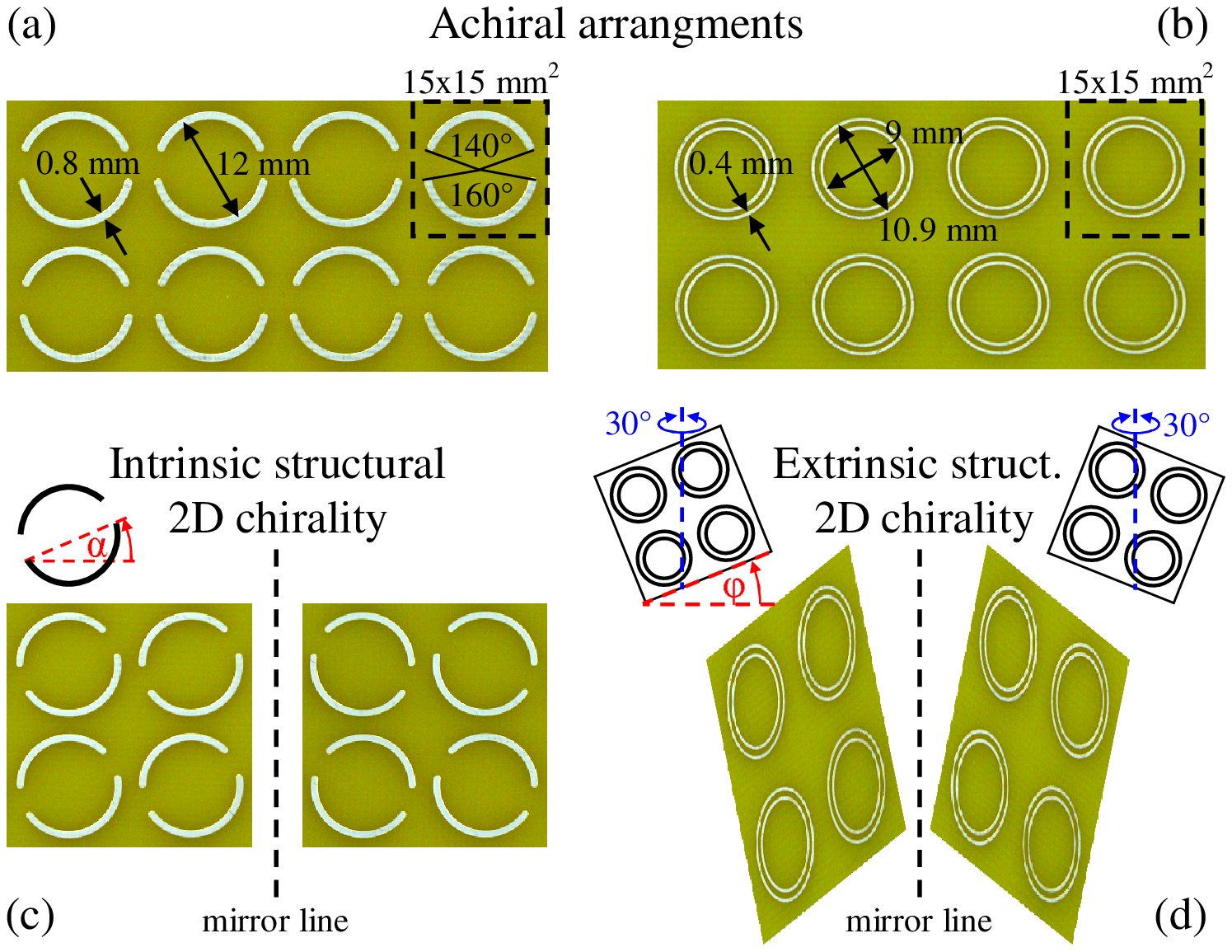}
\caption{\label{fig-chiral-array}\textbf{Metamaterial samples.}
Panels (a) and (b) show achiral arrangements of asymmetrically split
rings and double rings respectively. (c) Rotation of the split rings
by an angle $\alpha\neq n\cdot 45^\circ$, $n\in \mathbb{Z}$ leads to
intrinsic chirality of the array, which becomes different from its
mirror image. (d) At oblique incidence, even a regular array of
rings can become extrinsically 2D-chiral: For orientations
$\varphi\neq n\cdot 45^\circ$, $n\in \mathbb{Z}$ the projection of
the double ring array onto the plane normal to the direction of
incidence is 2D-chiral.}
\end{figure*}


In the experiments reported here we studied two different types of
planar metamaterials based on asymmetrically split rings and pairs
of concentric rings respectively, with the dimensions specified in
figure~\ref{fig-chiral-array}. Both metamaterial structures were
formed by a square array of about 200 meta-molecules separated by
15~mm, which rendered our structures non-diffracting below 13~GHz
for angles of incidence of up to $30^\circ$. The patterns were
etched on 1.6~mm thick lossy FR4 printed circuit boards
($\textrm{Im}~\varepsilon \sim 0.1$) covered with a
$35~\mu\textrm{m}$ copper layer using standard photolithography. The
transmission properties of the metamaterials were measured between 5
and 12~GHz in a microwave anechoic chamber using a vector network
analyzer (Agilent E8364B) and linearly polarized broadband horn
antennas (Schwarzbeck BBHA 9120D) equipped with lens concentrators.
In particular, we measured the structures' transmission matrix
$E^{0}_{i}=t_{ij}E^{t}_{j}$. To study chirality-related effects the
matrix was transformed to the circular polarization basis, where
indices $i$ and $j$ denote the handedness of the circularly
polarized components: ``+" for right-handed (RCP) and ``-" for
left-handed (LCP). In terms of power the transmission and
polarization conversion levels are given by $T_{ij}=|t_{ij}|^2$.

Eight different versions of the asymmetric split ring array were
studied at normal incidence, where the orientation of the split
$\alpha$ was varied in steps of $11.25^\circ=\pi/16~\textrm{rad}$
relative to the achiral arrangement shown in
figure~\ref{fig-chiral-array}(a). As illustrated in
figure~\ref{fig-chiral-array}(c), $\pm\alpha$ correspond to
structural planar chirality of opposite handedness, while rotations
by $\alpha$ and $\alpha+90^\circ$ yield identical metamaterial
arrays.

The double ring array was characterized at $30^\circ$ oblique
incidence for different orientations $\varphi$ of the array relative
to the plane of incidence. For $\varphi\neq n\cdot 45^\circ$ ($n\in
\mathbb{Z}$) the projection of the entire pattern onto the plane
normal to the propagation direction becomes 2D-chiral. Similarly to
the case of asymmetrically split rings, orientations $\pm\varphi$
correspond to extrinsically 2D-chiral arrangements of opposite
handedness, while an in-plane rotation of the metamaterial by
$90^\circ$ results in an identical experimental configuration.

\begin{figure*}[t!] \centering
\includegraphics[width=120mm]{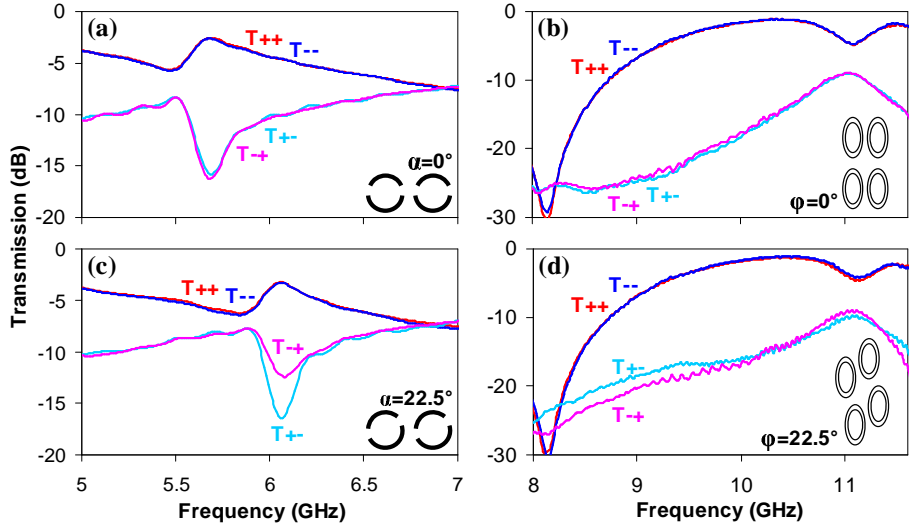}
\caption{\label{fig-transmission}Direct transmission $T_{++},
T_{--}$ and circular polarization conversion $T_{-+}, T_{+-}$
spectra for: (a) Normal incidence onto an achiral array of
asymmetrically split rings ($\alpha=0^\circ$). (b) $30^\circ$
oblique incidence onto the double ring array oriented in a way which
does not lead to extrinsic chirality ($\varphi=0^\circ$). (c) Normal
incidence onto an array of asymmetrically split rings which has
intrinsic structural planar chirality ($\alpha=22.5^\circ$). (d)
$30^\circ$ oblique incidence onto the double ring array rotated to
become extrinsically structurally 2D-chiral ($\varphi=22.5^\circ$).
Insets show the metamaterial patterns projected onto the plane
normal to the incident beam.}
\end{figure*}

Figure~\ref{fig-transmission} presents typical spectra of direct
transmission intensities $T_{++}, T_{--}$ and circular polarization
conversion $T_{-+}, T_{+-}$ for achiral and chiral arrangements of
both types of planar metamaterial \footnote{$T_{++}$ is the
co-coupling transmission intensity to a RCP wave from an incident
RCP wave, while $T_{-+}$ is the cross-coupling transmission
intensity to a LCP wave from an incident RCP wave.}. In all studied
cases, the direct transmission intensities (as well as field
transmission coefficients $t_{++}$ and $t_{--}$) did not depend on
the handedness or propagation direction of incident circularly
polarized waves. In particular this shows that in our case any 3D
chirality introduced by presence of the substrate
\cite{PRL_KuwataGonokami_2005_GiantOpticalActivity} is too small to
lead to significant optical activity, that is circular birefringence
$\textrm{arg}(t_{++})-\textrm{arg}(t_{--})$ or circular dichroism
$T_{++}-T_{--}$. The presence of circular polarization conversion
indicates a linearly birefringent / dichroic metamaterial response
in all cases. Figures~\ref{fig-transmission}(a) and
\ref{fig-transmission}(b) show that for both arrays of split rings
and double rings in the absence of structural 2D chirality the
intensities of circular polarization conversion are identical and
independent of the propagation direction, indicating the complete
absence of asymmetric transmission. When, however, the split rings
were rotated by an angle $\alpha$ to form a structurally 2D-chiral
array, normal incidence transmission through the metamaterial showed
a resonant region around 6~GHz where the right-to-left and
left-to-right polarization conversion efficiencies were different
from each other, $T_{-+}\neq T_{+-}$, as illustrated in
figure~\ref{fig-transmission}(c) for $\alpha = 22.5^\circ$. Here,
resonant excitation of the metamaterial occurs, when the effective
wavelength is twice as large as the average arc length
\cite{TrappedMode}. Similarly, when the double ring array was
rotated in its plane by an angle $\varphi$, so that for oblique
incidence its projection onto the plane normal to the wave
propagation direction became 2D-chiral, a broad band of asymmetric
circular polarization conversion appeared [as shown in
figure~\ref{fig-transmission}(d) for $30^\circ$ incidence and
$\varphi = 22.5^\circ$]. We note that the pronounced transmission
resonance at 8~GHz is associated with an electric dipole excitation
occurring when the length of the inner ring corresponds to the
effective wavelength \cite{DoubleRing}, however, in this case the
transmission asymmetry is not a resonant phenomenon. Our data
indicate that in both cases the conversion efficiencies are simply
interchanged for opposite directions of propagation,
$\overrightarrow{T}_{ij}= \overleftarrow{T}_{ji}$ and thus, for
example, RCP waves incident on front and back of the metamaterials
will experience different levels of circular polarization
conversion, $\overrightarrow{T}_{-+}\neq \overleftarrow{T}_{-+}$.
Given that the direct transmission terms are independent of the
propagation direction, the total transmission (i.e. transmission
measured with a polarization insensitive detector)
$T_+=T_{++}+T_{-+}$ is different for the circularly polarized waves
incident on front and back of the metamaterial arrays.

\begin{figure*}[t!] \centering
\includegraphics[width=170mm]{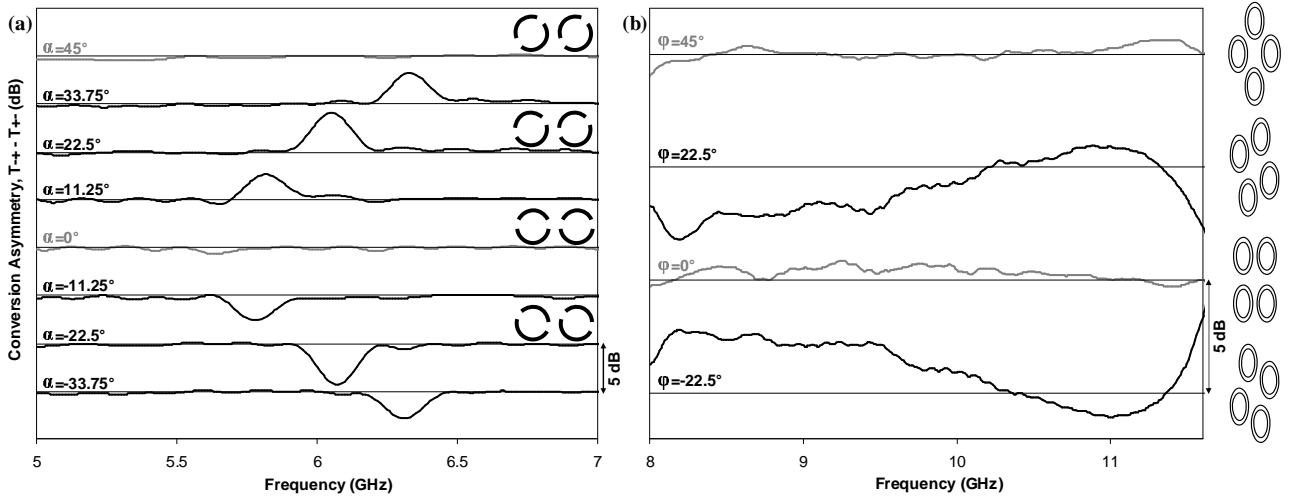}
\caption{\label{fig-asymmetry}(a) Conversion asymmetry,
$T_{-+}-T_{+-}$, for normal incidence onto arrays of asymmetrically
split rings as a function of the split's orientation $\alpha$. (b)
Conversion asymmetry for $30^\circ$ oblique incidence onto the
double ring metamaterial as a function of its in-plane orientation
$\varphi$. Insets show the metamaterial patterns as seen by an
observer looking along the incident beam.}
\end{figure*}

Figure~\ref{fig-asymmetry} illustrates the dependance of the
asymmetric effect on the arrangement of the meta-molecules in both
types of planar metamaterial array. When the split rings were
rotated by a multiple of $45^\circ$, structural 2D chirality of the
array was absent and asymmetric transmission could not be detected.
Other orientations of the split rings used in our experiments led to
a 0.2~GHz wide band of asymmetric transmission observed between 5.5
and 6.5~GHz [see figure~\ref{fig-asymmetry}(a)], whose exact
spectral position and magnitude were controlled by the split's
orientation $\alpha$. The largest asymmetry was observed for
$\alpha=\pm 22.5^\circ=\pm \pi/8~\textrm{rad}$, where the difference
in circular polarization conversion $T_{-+}-T_{+-}$ was about 4~dB.
As should be expected, mirror-forms $\pm \alpha$ of the split ring
array show asymmetric transmission of opposite sign [see
figure~\ref{fig-asymmetry}(a)]. The double ring array was
extrinsically 2D-chiral at oblique incidence for in-plane
orientations $\varphi$ excluding multiples of $45^\circ$ and, as
figure~\ref{fig-asymmetry}(b) illustrates, exhibited relatively wide
bands of asymmetric transmission with the sign of the effect being
reversed for enantiomeric forms of the array's projection.

Thus, our data confirm the existence of asymmetric transmission due
to both intrinsic and extrinsic structural chirality in arrays of
achiral meta-molecules. While asymmetric transmission was initially
only known for normal incidence onto lossy arrays of anisotropic and
intrinsically 2D-chiral meta-molecules
\cite{PRL_Fedotov_2006_AsymmetricTransmissionMW}, we are now able to
identify a much larger class of structures exhibiting the effect:
\emph{asymmetric transmission can occur for any lossy array of
particles, when its projection onto the plane normal to the
direction of incidence is 2D-chiral and anisotropic}. At oblique
incidence any regular array of even perfectly symmetric particles
can become planar chiral and anisotropic in projection, while at
normal incidence - when the structure and its projection coincide -
planar chirality and anisotropy must be properties of the array
itself.

However, our findings have much more far-reaching implications: The
observation of a signature 2D-chiral phenomenon (asymmetric
transmission) implies that also other effects connected to 2D
chirality should be observable in arrays of achiral building blocks.
For example 2D-chiral patterns show polarization rotation in
diffracted beams
\cite{PRL_Papakostas_2003_2dChirality,PRE_Prosvirnin_2003_chiralDiffraction}.
Our results imply that this 2D-chiral diffraction effect may be
possible at any regularly patterned interface. Furthermore, optical
activity in the form of circular birefringence
\cite{PRL_KuwataGonokami_2005_GiantOpticalActivity,
PRA_2DonSubstrateTheory, KuwataWaveguide} and circular dichroism
\cite{OptLett_Decker_2007_CircDichroismAlignedRosettes} as well as
circularly polarized second harmonic generation
\cite{L_NanoParticleSHG} have been observed at non-diffracting
planar chiral patterns on dielectric substrates making 3D-chiral
objects. Also patterns with extrinsic 2D chirality will become
3D-chiral if they are placed on a substrate or at the interface
between two different media. Therefore we may expect that even a
simple square array of spherical particles on a substrate could also
exhibit optical activity and circularly polarized second harmonic
generation at oblique incidence. Finally, as extrinsic 2D chirality
can arise from interaction of any directed quantity with a regular
pattern, chiral effects could even be envisaged for, e.g., a beam of
chiral molecules interacting with a regular achiral surface.

In summary, we have experimentally demonstrated that asymmetric
transmission can occur at highly symmetric periodically structured
interfaces. Our results imply that the effect may be expected for
oblique incidence onto any lossy periodically structured plane. Our
findings greatly expand the range of natural and artificial
materials in which the phenomenon may be expected, making asymmetric
transmission a mainstream electromagnetic effect rather than a
curiosity of planar chiral metamaterials. Indeed, while only few
natural examples of intrinsically 2D-chiral interfaces are known,
regular arrays of simple particles are much more common and much
easier to manufacture. This indicates that asymmetric transmission
should be observable in natural and self-assembled structures. Prime
candidates for asymmetric transmission at oblique incidence are
planar metamaterial structures, square arrays of plasmonic spheres
or semiconductor quantum dots and lossy double-periodic gratings,
which have the same symmetry, as the double ring array studied here.

\ack Financial support of the Engineering and Physical Sciences
Research Council, UK is acknowledged.

\section*{References}

\bibliographystyle{unsrt}

\end{document}